\documentclass[sigconf]{acmart}

\setcopyright{acmcopyright}
\copyrightyear{2019}
\acmYear{2019}
\acmDOI{10.1145/1122445.1122456}

\acmConference[ACM NanoCom 2019]{ACM NanoCom 2019: 6th ACM International Conference on Nanoscale Computing and Communication}{September 25--27, 2019}{Dublin, Ireland}
\acmBooktitle{ACM NanoCom 2019: 6th ACM International Conference on Nanoscale Computing and Communication,
    September 25--27, 2019, Dublin, Ireland}
\acmPrice{15.00}
\acmISBN{978-1-4503-9999-9/18/06}

\usepackage{amsmath,amssymb,amsfonts}
\usepackage{subcaption}
\DeclareCaptionFont{customCap}{\bfseries\boldmath}
\usepackage{circuitikz}
\usepackage{pgfplots}
\pgfplotsset{compat=1.14}
\usepackage{overpic}

\usepackage{siunitx}
\usepackage{xcolor}

\begin{document}

\title[Novel Receiver for SPIONs in an MC Setting]%
      {Novel Receiver for Superparamagnetic Iron Oxide Nanoparticles in a Molecular Communication Setting}

\author{Max Bartunik}
\authornote{Both authors contributed equally to this research.}
\email{max.bartunik@fau.de}
\author{Maximilian L{\"u}bke}
\authornotemark[1]
\email{maximilian.luebke@fau.de}
\affiliation{%
    \department{Institute for Electronics Engineering}
    \institution{Friedrich-Alexander-University Erlangen-Nuernberg (FAU)}
    \streetaddress{Cauerstr.~9}
    \postcode{91\,058}
    \city{Erlangen}
    \country{Germany}
}

\author{Harald Unterweger}
%\email{harald.unterweger@uk-erlangen.de}
\author{Christoph Alexiou}
%\email{christoph.alexiou@uk-erlangen.de}
\affiliation{%
    \department{Section of Experimental Oncology and Nanomedicine (SEON)}
    \institution{University Hospital Erlangen}
    \streetaddress{Gl{\"u}ckstr.10a}
    \postcode{91\,054}
    \city{Erlangen}
    \country{Germany}
}

\author{Sebastian Meyer}
%\email{sebastian.meyer@fau.de}
\author{Doaa Ahmed}
%\email{doaa.ahmed@fau.de}
\author{Georg Fischer}
%\email{georg.fischer@fau.de}
\affiliation{%
    \department{Institute for Electronics Engineering}
    \institution{FAU, Erlangen, Germany}
%    \streetaddress{Cauerstr.~9}
%    \postcode{91\,058}
%    \city{Erlangen}
%    \country{Germany}
}

\author{Wayan Wicke}
%\email{wayan.wicke@fau.de}
\author{Vahid Jamali Kooshkghazi}
%\email{vahid.jamali@fau.de}
\author{Robert Schober}
%\email{robert.schober@fau.de}
\affiliation{%
    \department{Institute for Digital Communication}
    \institution{FAU, Erlangen, Germany}
%    \streetaddress{Cauerstr.~7}
%    \postcode{91\,058}
%    \city{Erlangen}
%    \country{Germany}
}

\author{Jens Kirchner}
\orcid{0000-0002-8623-9551}
\email{jens.kirchner@fau.de}
\affiliation{%
    \department{Institute for Electronics Engineering}
    \institution{FAU, Erlangen, Germany}
%    \streetaddress{Cauerstr.~9}
%    \postcode{91\,058}
%    \city{Erlangen}
%    \country{Germany}
}

\renewcommand{\shortauthors}{Bartunik and L{\"u}bke, et al.}

\begin{abstract}
Superparamagnetic iron oxide nanoparticles (SPIONs) have recently been introduced as information carriers in a testbed for molecular communication (MC) in duct flow. Here, a new receiver for this testbed is presented, based on the concept of a bridge circuit.
The capability for a reliable transmission using the testbed and detection of the proposed receiver was evaluated by sending a text message and a \SI{80}{bit} random sequence at a bit rate of \SI{1}{\per\s}, which resulted in a bit error rate of \SI{0}{\percent}. Furthermore, the sensitivity of the device was assessed by a dilution series, which gave a limit for the detectability of peaks between \SIrange{0.1}{0.5}{\milli\gram\per\milli\liter}.
Compared to the commercial susceptometer that was previously used as receiver, the new detector provides an increased sampling rate of \SI{100}{samples\per\s} and flexibility in the dimensions of the propagation channel. Furthermore, it allows to implement both single-ended and differential signaling in SPION-bases MC testbeds.
\end{abstract}

%\begin{CCSXML}
%    <ccs2012>
%    <concept>
%    <concept_id>10010583.10010584.10010587</concept_id>
%    <concept_desc>Hardware~PCB design and layout</concept_desc>
%    <concept_significance>500</concept_significance>
%    </concept>
%    <concept>
%    <concept_id>10010583.10010588.10010596</concept_id>
%    <concept_desc>Hardware~Sensor devices and platforms</concept_desc>
%    <concept_significance>500</concept_significance>
%    </concept>
%    <concept>
%    <concept_id>10010583.10010588.10010595</concept_id>
%    <concept_desc>Hardware~Sensor applications and deployments</concept_desc>
%    <concept_significance>300</concept_significance>
%    </concept>
%    </ccs2012>
%\end{CCSXML}
%
%\ccsdesc[500]{Hardware~PCB design and layout}
%\ccsdesc[500]{Hardware~Sensor devices and platforms}
%\ccsdesc[300]{Hardware~Sensor applications and deployments}

%    \includegraphics[width=\textwidth]{sampleteaser}
%    \caption{Seattle Mariners at Spring Training, 2010.}
%    \Description{Enjoying the baseball game from the third-base
%        seats. Ichiro Suzuki preparing to bat.}
%    \label{fig:teaser}

\maketitle

\section{Introduction}
In the exploration of molecular communication strategies (see \cite{Nakano2013, Jamali:2019} for an overview), testbeds play a central role as they allow to evaluate communication theory, to identify new physical aspects that have to be taken into account such as sources of interference and to provide steps towards implementation of molecular communication for practical systems applications.

For that purpose, testbeds have been proposed based on alcohol
\cite{Farsad2013, Koo2016}
(see also \cite{Wang2015} for spatial instead of temporal coding)
and acids/bases
\cite{Farsad2017}
(refer to \cite{Grebenstein2019, Krishnaswamy2013} for corresponding transmitter and receiver designs)
as signaling molecules.
A third type of information carrier was proposed by the authors in \cite{Unterweger2018} for a testbed based on superparamagnetic iron oxide nanoparticles (SPIONs). In contrast to alcohol and acids/bases, these particles, which were originally developed for magnetic drug delivery, are biocompatible and hence applicable for use in humans. They can be fabricated relatively easily and are detected as a change in inductance of a measurement coil wound around the propagation channel. Hence, detectors can be operated noninvasively, i.\,e., they do not have to be inserted into the propagation channel, which provides a major advantage for human use, e.\,g., with blood vessels as propagation channel.

However, a major challenge when implementing testbeds for MC is the lack of appropriate detectors. In the present case, the receiver that was used in \cite{Unterweger2018}, the commercial susceptometer Bartington\textsuperscript{\textregistered} MS2G, was not operated according to its original measurement purpose, i.\,e., the characterization of material samples. It therefore exhibited two considerable disadvantages:
First, the detector coil had a fixed width. This, on the one hand, posed restrictions on the maximum width of the propagation channel. On the other hand, for channel widths smaller than the detector width, enhancement of the measurement signal and thus of the detector sensitivity by reducing the width of the detector coil was not possible. Second, the susceptometer, which is not intended for dynamic measurements, provided a maximum sample rate of only \SI{10}{samples\per\s}. Hence, in order to achieve higher flexibility in the dimensions of the propagation channel and to increase both the temporal and the measurement resolution, a new customized detector device for SPION-based MC is needed. In this paper, a first prototype of this device is presented.

The proposed device offers yet another feature: It incorporates a second coil to provide a reference signal. Hence, the detector can not only be operated in a single-ended signaling design with one propagation channel as in previous testbeds for molecular communication, but can also be used for a differential signaling design, where information is encoded in the difference of SPIONs concentration between two tubes that serve as propagation channel.

In the following sections, the testbed will be outlined (for more details, see \cite{Unterweger2018}) and the design of the proposed detector will be described. The device is evaluated by transmitting an exemplary text message and by assessing its sensitivity for different concentrations of SPION solutions. A discussion of the achievements and potential future applications concludes the article.
\section{Testbed}
\label{sec:channelmodel}
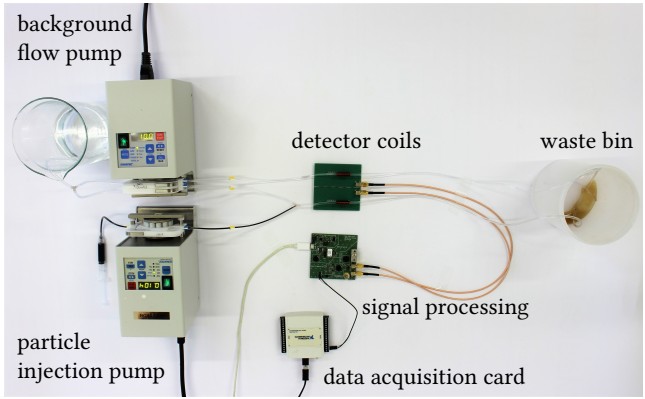
\begin{figure}
    \centering
    \begin{overpic}[width=\columnwidth, angle=180]{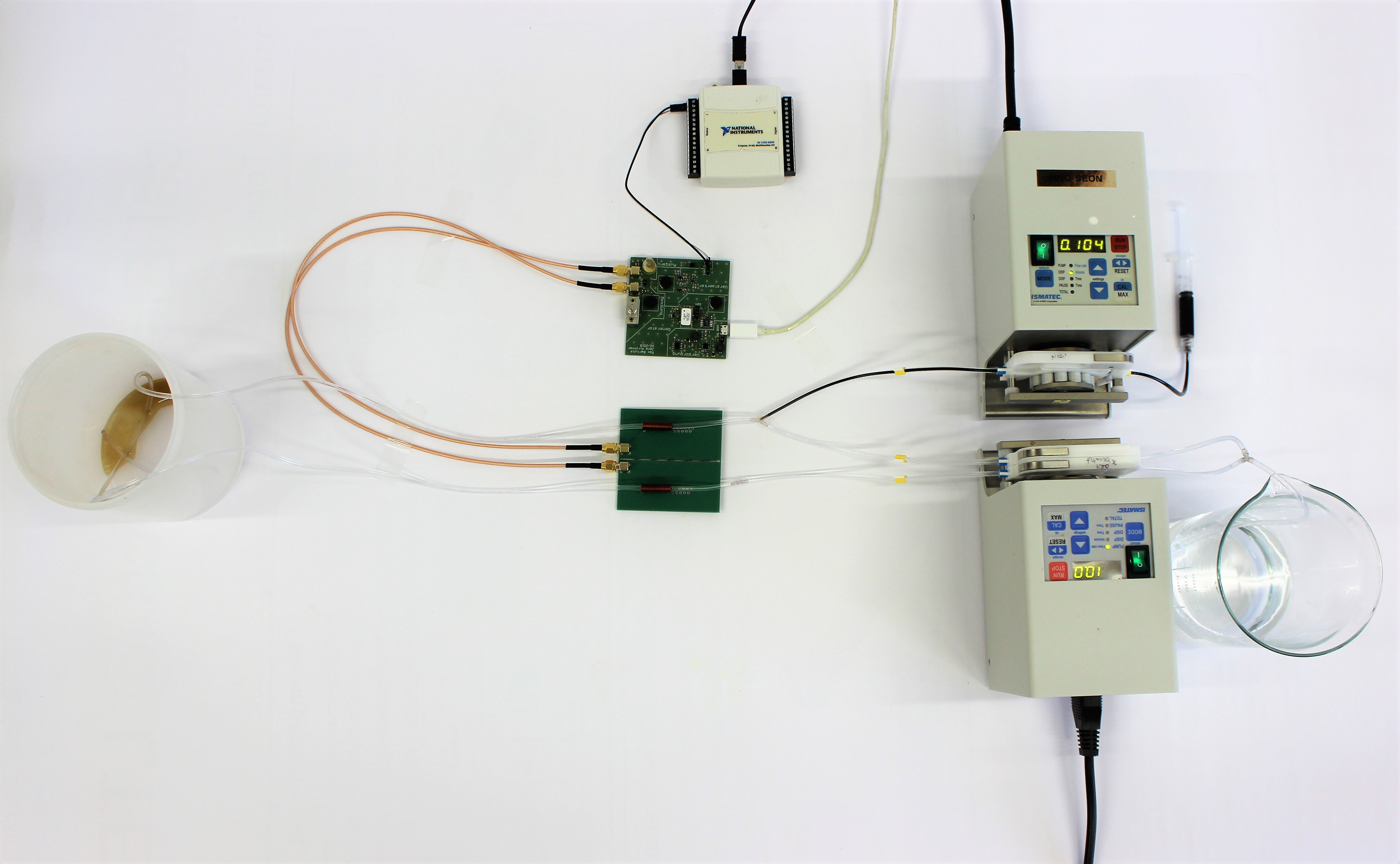}
        \put (2, 5){\parbox{100pt}{particle\\ injection pump}}
        \put (2, 55){\parbox{100pt}{background\\ flow pump}}
        \put (45, 39){detector coils}
        \put (56, 13){signal processing}
        \put (50, 2){data acquisition card}
        \put (84, 39){waste bin}
    \end{overpic}
    \caption{Testbed. From left to right are shown:
        Transmitter consisting of water reservoir for background flow,
        syringe as reservoir for particle injection,
        two pumps for providing the background flow and the particle injection
        (upper and lower pump, respectively)
        and Y-connector;
        propagation channel after the Y-connetor;
        receiver consisting of the PCB's for detector coils and signal processing
        (upper and lower PCB, respectively),
        data acquisition card and waste bin.}
    \label{fig: Testbed}
\end{figure}

Figure \ref{fig: Testbed} shows the complete testbed including the propagation channel.

\subsection{Information Carriers}
Information is encoded in the SPION concentration in aqueous solution: A binary ``1'' is represented by the injection of SPION particles into the propagation channel, a binary ``0'' by no particles being released.

SPIONs originate from biomedical applications and thus have an established biocompatibility with ongoing studies on biosafe applications \cite{Lindemann2015}. The nanoparticles used as transmitter molecules in this setup where synthesized by the Section of Experimental Oncology and Nanomedicine (SEON) of the University Hospital Erlangen. They have a hydrodynamic radius of \SI{50}{\nano\m} and a susceptibility of \num{8.78d-3} for \SI{1}{\milli\gram\per\milli\liter}. They are suspended in water with an estimated particle concentration of \SI{5d13}{particles\per\milli\liter} for a stock concentration of \SI{10}{\milli\gram Fe\per\milli\liter}.

\subsection{Transmitter}
The transmitter consists of the following components:
a reservoir with distilled water, which together with a first pump (Ismatec\textsuperscript{\textregistered} ISM831C) provides the channel medium in form of a background flow with a continuous rate of \SI{10}{\milli\liter\per\min};
a syringe as reservoir for the SPION suspension;
a second, computer controlled pump (Ismatec\textsuperscript{\textregistered} ISM596D) that injects a defined amount of this SPION suspension into the transmission channel. For that purpose, the tubes for background flow and particle injection are joint via a Y-connector. The pump for SPION injection is controlled with a LabView application that allows to encode the desired binary sequence in a series of particle releases, each with a volume of \SI{104}{\micro\liter} and at a flow rate of \SI{10}{\milli\liter\per\min}. A symbol duration of  \SI{1}{\s} was employed.

\subsection{Propagation Channel}
The propagation channel consists of a tube with a diameter of \SI{0.84}{\milli\m} and a length of \SI{5}{\centi\m}. The channel begins after the Y-connector that joins background flow and particle injection and ends at the detector coil of the receiver.

\subsection{Receiver}
The bursts of SPIONs traveling through the propagation channel are measured with a detector coil as the SPIONs change the inductance value of the coil while they pass through it. This basic principle is exploited in magnetic susceptometers \cite{Das2013}. In this paper, the commercial susceptometer that was used in previous studies \cite{Unterweger2018} (Bartington MS2G) is replaced by the device described in the following section.
\section{Device Conception and Realization}
As in the previously used, commercial susceptometer, the nanoparticle are detected by a change of inductance of a coil that they pass through. For the proposed design, two coils within a bridge circuit are utilized, one as reference and one for measurement \cite[ch.~32]{Williams:2011}. Combined with an appropriate capacitor each coil forms a resonator circuit that is driven by an amplified generator signal. SPIONs passing through the measurement coil cause a change in inductance and as a consequence also a shift in resonance frequency. As a result the bridge circuit is unbalanced and the resulting differential voltage is detected via an instrumentation amplifier and digital processing.

\subsection{Power Supply}
To amplify the generator signal and for use in the instrumentation amplifier, operational amplifiers that are driven by a differential power source ($\pm\SI{5}{\volt}$ and $\pm\SI{15}{\volt} $) were selected. To simplify the external power supply, all required voltages are produced by the device from a single \SI{5}{\volt} source.

To supply the required $\pm\SI{15}{\volt}$ voltage from the \SI{5}{\volt} source, the split-rail converter TPS65131 (Texas Instruments) was used. The \SI{-5}{\volt} rail was provided by reducing the \SI{-15}{\volt} output via a negative voltage regulator (LM79L05, Texas Instruments).

\subsection{Signal Generator}
The resonator circuit is driven by a sine wave source with a frequency of \SI{10}{\mega\hertz}, namely the ultra-low phase noise sine wave generator CVSS-945-10000 by Crystek. The voltage-tunable output frequency was set to \SI{10}{\mega\hertz} by means of a \SI{2.5}{\volt} low-dropout regulator (LP5907, Texas Instruments). A load of \SI{50}{\ohm} was connected to the signal output, resulting in a peak-to-peak voltage of \SI{3}{\volt}.

As a higher voltage swing is desired, a current feedback amplifier (LT1223, Linear Technology)  with a large output drive and a gain bandwidth of \SI{100}{\mega\hertz} was used to amplify the generated signal to a peak-to-peak voltage of approximately \SI{15}{\volt}. As the optimal amplification factor was determined during testing, a potentiometer was fitted to adjust the resulting output voltage.

\subsection{Bridge Circuit}
\begin{figure}
	\centering
	\begin{circuitikz}[scale=0.55, /tikz/circuitikz/bipoles/length=8mm]
		\draw
		(-3, 14) to [vsourcesin, l=\SI{10}{\mega\hertz}] (-3, 21) -- (1, 21) -- (7,21)
		(-3,14) node[ground] {}

		(1,21) -- (1,20) to [resistor, l=$R_1$] (1,18) to [short, *-] (1,17)
		(1,17) -- (-1,17) to [variable capacitor, l=$C_{11}$] (-1,15) -- (1,15)
		(1,17) -- (1,17) to [capacitor, l=$C_{12}$] (1,15) -- (1,15)
		(1,17) -- (3,17) to [inductor, l=$L_1$] (3,15) -- (1,15)
		(1,15) -- (1,14) -- (4,14)

		(7,21) -- (7,20) to [potentiometer, l=$R_2$] (7,18) to [short, *-] (7,17)
		(7,17) -- (5,17) to [variable capacitor, l=$C_{21}$] (5,15) -- (7, 15)
		(7,17) -- (7,17) to [capacitor, l=$C_{22}$] (7,15) -- (7, 15)
		(7,17) -- (9,17) to [inductor, l=$L_2$] (9,15) -- (7,15)
		(7,15) -- (7,14) -- (4,14)

		(4,14) node[ground]{}

		(1,18) to [open, v^=$V_\text{diff}$] (7,18)
		;
	\end{circuitikz}
	\caption{Bridge circuit.
             The tunable capacitances $C_{11}$, $C_{21}$ and tunable resistance $R_2$
             are used to balance the bridge during calibration. SPIONs that flow through
             the detector coil change its inductance (either $L_1$ or $L_2$), which
             results in a non-zero voltage difference $V_\text{diff}$.}
    \label{fig:wheatstonecircuit}
\end{figure}
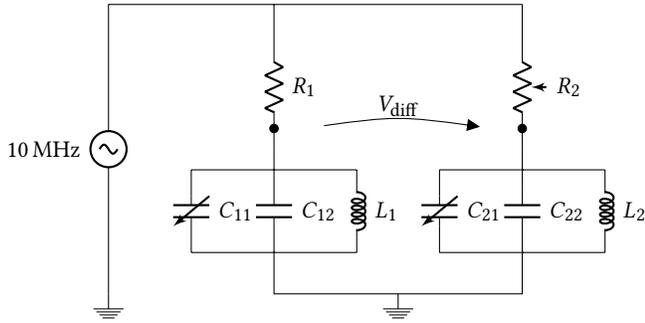

A bridge circuit, typically consisting of four impedances, is implemented here with two resistors and the resonator circuits for the detector coils as shown in Fig.~\ref{fig:wheatstonecircuit}. A potentiometer is used as resistor $R_2$ and both branches contain adjustable capacitances ($C_{11}$ and $C_{21}$) to allow tuning of the resonance frequency. Simple switching of detector coils is enabled by use of SMA- (SubMiniature version A) headers.

Ideally, if correctly tuned, the voltage $V_\text{diff}$ between the two branches of the bridge circuit is zero as long as no nanoparticles pass through the detector coils. As soon as a difference in inductance of one coil (either $L_1$ or $L_2$) occurs the corresponding resonator circuit will be de-tuned, the impedance changes, and the branches of the bridge become unbalanced. 

By tuning the branches of the bridge circuit equally, a peak-to-peak difference voltage $V_\text{diff}$ below \SI{40}{\milli\volt} could by achieved. This value hence constitutes the limit of tunability of the two branches by use of the tunable resistor and capacitances.

\subsection{Detector Coils}
To fit the requirements and geometry of the setup ideally, special detector coils were produced. They have an inner diameter that allows a tube with a girth of up to \SI{3.5}{\milli\m} to be exchanged easily without a needlessly large air gap remaining. To this end, bondable enamelled wire with a diameter of \SI{1}{\milli\m} was wound around a steel rod (\SI{4}{\milli\m}) to produce a coil of approximately \SI{20}{\milli\m} length with 20 windings. The coil was heated thoroughly under a hot air flow to bond the individual windings, ensuring mechanical stability and therefore a constant inductance value.

\begin{figure}
	\centering
	\includegraphics[width=0.4\textwidth]{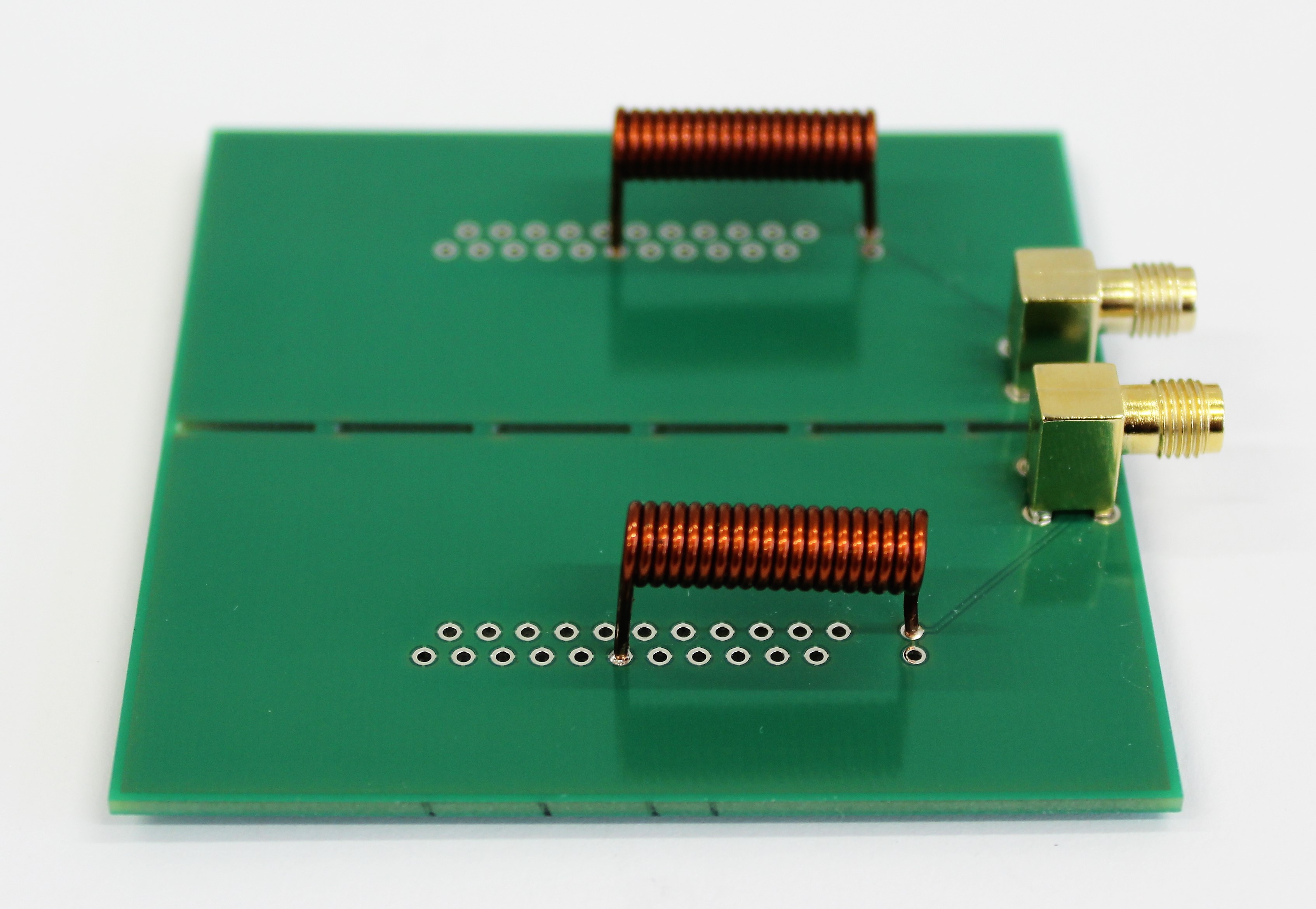}
	\caption{Detector coils mounted on PCB}
    \label{fig:coils}
\end{figure}

For reasons of usability the produced coils were mounted on a printed circuit board (PCB) and attached to SMA headers (see Fig.~\ref{fig:coils}).

\subsection{Instrumentation Amplifier}
To simplify the detection of voltage level changes in the bridge circuit, an instrumentation amplifier was implemented using two current feedback amplifiers of the type LT1395 from Linear Technology as input stage and the operation amplifier AD8033 from Analog Devices as amplification stage. The chosen devices show low noise over a sufficient bandwidth. The differential amplification of the instrumentation amplifier can be adjusted with a potentiometer.

The maximal amplification factor in the real-life application was limited by the current the power source could provide. After tuning the branches of the bridge circuit, the amplification of the instrumentation amplifier was adjusted to an output peak-to-peak voltage of \SI{1}{\volt}.

\subsection{Digital Measurement}
A simple digital measurement is made possible through an envelope detector connected to the output of the instrumentation amplifier. The result is a DC-signal that varies in the range of \SIrange{100}{500}{\milli\volt}. The output signal was captured with the data acquisition module NI USB-6009 from National Instruments. The module has a resolution of \SI{11}{bit} with a programmable-gain amplifier and a maximal sample rate of \SI{d4}{samples\per\s}. The supplied LabView application ``NI-DAQmx Base Data Logger'' was used to digitally record measured values. A sample rate of \SI{100}{samples\per\s} was used for all measurements.

\subsection{Printed Circuit Board}
\begin{figure}
    %\begin{subfigure}{0.48\textwidth}
    %    \centering
    %    \includegraphics[height=0.8\columnwidth, trim=95 0 100 0, clip]{PCB_Wheatstone}
    %    \caption{PCB layout (without ground planes).}
    %    \label{fig:pcb}
    %\end{subfigure}
    %\hfill
    %\begin{subfigure}{0.48\textwidth}
    	\centering
    	\includegraphics[height=0.7\columnwidth, trim=50 50 20 50, clip]{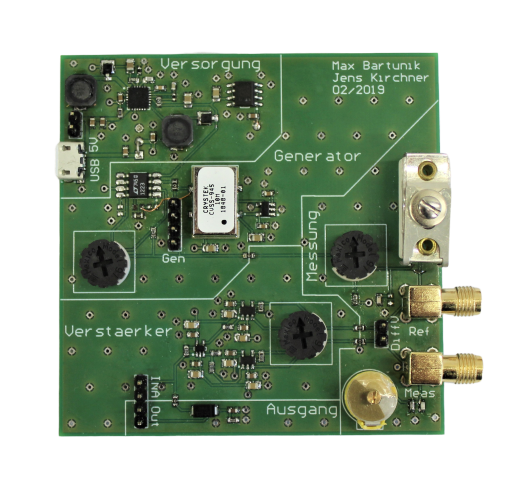}
   % 	\caption{Assembled detection device.}
    %    \label{fig:dev}
    %\end{subfigure}
    \caption{Signal acquisition circuitry.}
    \label{fig: Device: PCB}
\end{figure}

The described circuitry was implemented on a four layered PCB consisting of \SI{1.55}{\milli\m} thick FR-4 composite material with a square layout of \SI{75}{\milli\m} by \SI{75}{\milli\m}. %as shown in Fig.~\ref{fig: Device: PCB}(a).
Layers two and four, consisting solely of ground planes, act as shielding. Layer 1 contains the signal components, while the power distribution is implemented in layer 3. All remaining space is covered with ground planes and connected through a mesh of vias. Additionally, a via is placed closed to every surface mounted device (SMD) with a ground connection. Figure \ref{fig: Device: PCB} shows the completely assembled detector device.

Altium Designer 18.1.7 was used to layout the PCB, which was produced by the Multi Leiterplatten GmbH, Germany, and assembled with the own department facilities.

\subsection{Tuning Procedure}
Balancing the bridge circuit is an important step for the functionality of the device. To maximize the sensitivity as SPIONs pass through the detector coil, the resonator circuits should ideally be in resonance at \SI{10}{\mega\hertz} and identical to each other. Due to small differences in the inductance of the detector coils, a trade-off between ideal resonance and identical resonance must be found.

In the first step, the PCB was fitted with SMD capacitances ($C_{12}$ and $C_{22}$ in Fig.~\ref{fig:wheatstonecircuit}) that shifted the resonance frequency into the desired range. The tunable capacitances $C_{11}$ and $C_{21}$ were then adjusted to maximize the signal output of each channel and therefore maximize resonance. Using the potentiometer $R_2$, the peak-to-peak voltage of both channels was matched. Finally, to compensate for the influence of the used measurement tips, with a given capacitance of \SI{16}{\pico\farad}, the tuning steps were repeated while measuring the output of the instrumentation amplifier, which in turn was adjusted to a peak-to-peak voltage of \SI{1}{\volt}. All measurements were performed with the oscilloscope WaveRunner LT262 (LeCroy).

\section{Device Evaluation}
%\section{Coding Scheme}
The proposed measurement device was evaluated in two respects:
As a proof of concept, a simple text message was sent with the testbed. Furthermore, the sensitivity of the device was assessed by comparing pulse series with different concentrations of SPION suspension.

\subsection{Text Message Transmission and Long Random Binary Sequence}
A simple coding table was devised to transmit text messages composed of capital letters efficiently. Each codeword is headed by a logical ``1'' to allow receiver synchronization, followed by 5 bit representing the character to be coded. This latter code component is the 5-bit binary representation of the number of the letter within the Latin alphabet, starting with zero for letter ``A''. Hence,
\begin{align*}
    \text{``A''} &\equiv \text{``100000''},\quad 
    \text{``B''}  \equiv \text{``100001''},\quad... \\
    \text{``F''} &\equiv \text{``100101''},\quad... \\
    \text{``U''} &\equiv \text{``110100''},\quad...
\end{align*}

With this coding scheme, the sample sequence ``FAU'' with binary representation
\begin{equation}
    100101\,100000\,110100
    \label{eq: Results: FAU binary sequence}
\end{equation}
was transmitted using a suspension with a particle concentration of \SI{7.5}{\milli\gram\per\milli\liter}.

At the receiver, peak detection was performed by use of Matlab (MathWorks), before decoding the bit sequence.
First noise in the received data was reduced by applying a moving average filter with a width of 21 samples. Next, a threshold to differentiate between the logical values ``1'' and ``0'' was calculated from the received data as the mean between the highest and the lowest value. Finally, the rising edges were identified as threshold-crossings with positive slope. Based on the time $t_0$ of the first detected rising edge and symbol interval $T_\text{S}$, symbol intervals were defined by (for interval no. $k$)
\begin{equation}
    I_k = \left[ t_0 + kT_\text{S}; t_0 + (k+1)T_\text{S} \right[\,.
\end{equation}
Bit no. $k$ was set to ``1'' if the threshold was exceeded for at least $30$\,\%  of samples in $I_k$, and to ``0'' otherwise.

A further test with a random sequence of \SI{80}{bit} was performed:
\begin{equation}
\begin{aligned}
    11001001\, 11111110\, 10110011\,01101000\,10001010 \\
    01011001\,01100011\,11111000\,11010101\,00001000
    \label{eq: Results: random binary sequence}
\end{aligned}
\end{equation}

\subsection{Sensitivity Test}
To characterize the quality of detection and allow for comparison to other devices, the sensitivity of the receiver was determined. To this end, the bit sequence ``10101'' was transmitted using various dilutions of the SPION suspension. Solutions with a particle concentration of \SI{10}{\milli\gram\per\milli\liter}, %\SI{7.5}{\milli\gram\per\milli\liter}, 
\SI{5.0}{\milli\gram\per\milli\liter},
\SI{1.0}{\milli\gram\per\milli\liter},
\SI{0.5}{\milli\gram\per\milli\liter} and
\SI{0.1}{\milli\gram\per\milli\liter} were tested.
\section{Results}

\subsection{Comparison}
\begin{figure}
        \begin{tikzpicture}
        \pgfplotsset{set layers}
        \begin{axis}[
            width=0.7\columnwidth, height=0.52\columnwidth,
            scale only axis,
            axis y line*=left, % the ’*’ avoids arrow heads
            ylabel=\textcolor{blue}{Measured signal (\si{\volt})},
            yticklabel style={blue}, ytick style={blue}, y axis line style={blue},
            xlabel = Time {(s)}, %ylabel = Susceptibility {[V]},
            xmin=0, xmax=8]
            \input{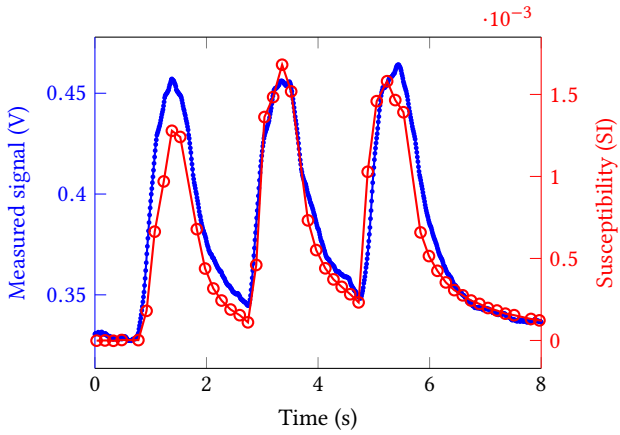}
        \end{axis}
        \begin{axis}[
            width=0.7\columnwidth, height=0.52\columnwidth,
            scale only axis,
            axis y line*=right,
            ylabel=\textcolor{red}{Susceptibility (SI)},
            ylabel near ticks, yticklabel pos=right,
            yticklabel style={red}, ytick style={red}, y axis line style={red},
            xmin=0, xmax=8,
            axis x line=none]
        \addplot [red, thick, mark=o] coordinates{
            (0.0299999999999976 , -0.0000013595)
            (0.179999999999996 , 0)
            (0.330999999999996 , -0.0000020111)
            (0.479999999999997 , 0.0000026701)
            (0.779 , 0.0000020111)
            (0.925999999999998 , 0.0001805253)
            (1.072 , 0.0006632854)
            (1.234 , 0.0009695049)
            (1.38 , 0.0012783)
            (1.529 , 0.0012406)
            (1.82699999999999 , 0.0006785344)
            (1.975 , 0.0004384555)
            (2.12299999999999 , 0.000316791)
            (2.27 , 0.0002433893)
            (2.432 , 0.0001884868)
            (2.59599999999999 , 0.0001540983)
            (2.743 , 0.0001104452)
            (2.892 , 0.0004602501)
            (3.039 , 0.0013622)
            (3.195 , 0.0014839)
            (3.356 , 0.001681)
            (3.518 , 0.0015183)
            (3.819 , 0.0007320623)
            (3.967 , 0.0005508705)
            (4.13 , 0.0004397699)
            (4.276 , 0.0003722886)
            (4.424 , 0.0003267184)
            (4.57699999999999 , 0.0002817359)
            (4.72699999999999 , 0.0002321248)
            (4.89 , 0.0010283)
            (5.052 , 0.0014588)
            (5.238 , 0.0015805)
            (5.384 , 0.0014654)
            (5.531 , 0.0013914)
            (5.837 , 0.0006586606)
            (5.988 , 0.0005144672)
            (6.14 , 0.0004245811)
            (6.29099999999999 , 0.0003537894)
            (6.442 , 0.0003068334)
            (6.593 , 0.0002751076)
            (6.744 , 0.0002433893)
            (6.89399999999999 , 0.0002228752)
            (7.046 , 0.0001977364)
            (7.207 , 0.0001825363)
            (7.35799999999999 , 0.0001633479)
            (7.515 , 0.0001540983)
            (7.817 , 0.0001309255)
            (7.968 , 0.0001223199)
            (8.124 , 0.0001176951)
            (8.28499999999999 , 0.0001104603)
            (8.436 , 0.0001018059)
            (8.587 , 0.0000971886)
            (8.738 , 0.0000879314)
            (8.89 , 0.0000813256)
            (9.04099999999999 , 0.0000760718)
            (9.192 , 0.000071447)
            (9.343 , 0.0000648074)
            (9.50099999999999 , 0.0000608454)
            (9.813 , 0.000050933)
            (9.973 , 0.0000475888)
            };
        \end{axis}
        \end{tikzpicture}
    \caption{Comparison of measured curves by use of the commercial susceptometer
             that was previously used (red) and the proposed detector device (blue).}
    \label{fig: Results: Comparison Suscept vs Proposition}
\end{figure}

Figure~\ref{fig: Results: Comparison Suscept vs Proposition}
gives a qualitative comparison of the measured receiver curves of the susceptometer that was used previously in \cite{Unterweger2018} and of the proposed device (in red and blue, respectively). The figure clearly demonstrates the increased resolution of the signal with the new detector.

Compared to the susceptometer, the sampling rate of the proposed detector of \SI{100}{samples\per\s} is already 10 times larger. It was chosen due to restrictions of the ready to use graphical user interface (GUI) that accompanied the data acquisition card. With a customized data acquisition software, the sampling rate can further be increased.

\subsection{Sequence transmission}
\begin{figure}
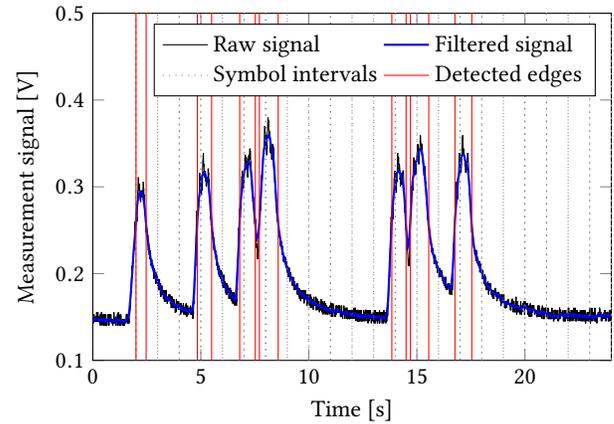

\centering
\begin{tikzpicture}
\begin{axis}[
    width=\columnwidth, height=6.2cm, %0.6\columnwidth,
	xlabel = Time {[s]},
	ylabel = Measurement signal {[V]},
	legend pos = north east,
	legend cell align ={left},
    ymin=0.1, ymax=0.5,
    xmin=0, xmax=24,
    legend columns=2]
	
	\input{addplot__TransmTest_Signal}
	\addlegendentry{Raw signal}
	
    \input{addplot__TransmTest_FiltSignal}
    \addlegendentry{Filtered signal}
    
    \input{addplot__TransmTest_Intervals}
    \addlegendentry{Symbol intervals}

    \input{addplot__TransmTest_EdgeDetection}
    \addlegendentry{Detected edges}
	
\end{axis}
\end{tikzpicture}
\caption{Transmission test with text sequence ``FAU''
        (binary representation see (\ref{eq: Results: FAU binary sequence})).
        Raw and filtered measurement signals are plotted in black and blue color, respectively.
        The symbol intervals, as derived from the first detected rising edge, are
        indicated by dotted gray vertical lines, detected edges by solid red lines.
        Sufficient values above the threshold are found in each symbol interval that
        corresponds to a binary ``1''.}
\label{plot:fau}
\end{figure}

The sample sequence ``FAU'' was successfully transmitted, as can be seen in Fig.~\ref{plot:fau}. All bits were detected correctly using the simple method described in the previous section. 

\begin{figure}
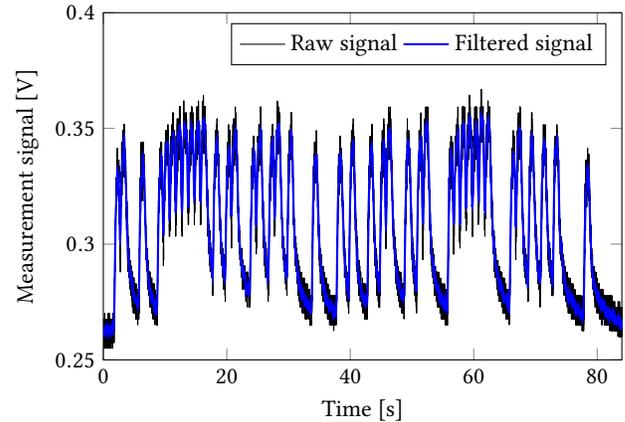

\centering
\begin{tikzpicture}
\begin{axis}[
    width=\columnwidth, height=6.2cm, %0.6\columnwidth,
	xlabel = Time {[s]},
	ylabel = Measurement signal {[V]},
	legend pos = north east,
	legend cell align ={left},
    ymin=0.25, ymax=0.4,
    xmin=0, xmax=84,
    legend columns=2]
	
	\input{addplot__RandomTest_Signal}
	\addlegendentry{Raw signal}
	
    \input{addplot__RandomTest_FiltSignal}
    \addlegendentry{Filtered signal}
    
\end{axis}
\end{tikzpicture}
\caption{Transmission test with \SI{80}{bit} random sequence
        (see (\ref{eq: Results: random binary sequence})).
        Raw and filtered measurement signals are plotted in black and blue color, respectively. Each peak corresponding to a binary ``1'' can be clearly identified.}
\label{plot:random}
\end{figure}

A second test with a random \SI{80}{bit} sequence was performed, again with the optimal result of all peaks corresponding to a ``1'' being correctly detected (see Fig.~\ref{plot:random}).

A characteristic waveform for each burst of particles with a quick rise time and a delayed return to zero, as the particles are washed out of the detector coil by the background flow, can be observed. Due to the slow decay of particles in the tube because of laminar flow, successive bursts cause an accumulation of remaining particles which increases the measured concentration and further delays the return to zero, i.\,e., inter-symbol interference occurs.

\subsection{Sensitivity Test}
\begin{figure*}
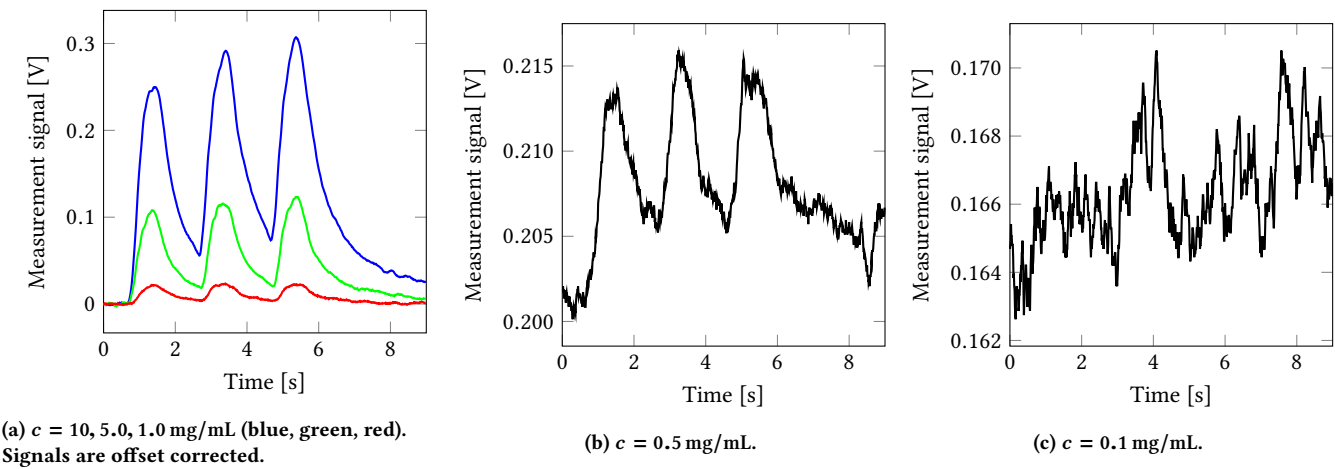

    \begin{subfigure}{0.33\textwidth}
        \centering
        \begin{tikzpicture}
        \begin{axis}[
            width=\textwidth, height=\textwidth,
            xlabel = Time {[s]}, ylabel = Measurement signal {[V]},
            xmin=0, xmax=9]
            \input{addplot__Dilution10__offsetcorr}
            \input{addplot__Dilution5__offsetcorr}
            \input{addplot__Dilution1__offsetcorr}
        \end{axis}
        \end{tikzpicture}
        \caption{$c = \SIlist{10; 5.0; 1.0}{\milli\gram\per\milli\liter}$
                 (blue, green, red). \\Signals are offset corrected.}
    \end{subfigure}
    \begin{subfigure}{0.33\textwidth}
        \centering
        \begin{tikzpicture}
        \begin{axis}[
            width=\textwidth, height=\textwidth,
            xlabel = Time {[s]}, ylabel = Measurement signal {[V]},
            xmin=0, xmax=9,
            y tick label style={
                /pgf/number format/.cd, fixed, fixed zerofill,
                precision=3, /tikz/.cd
            }]
            \input{addplot__Dilution05}
        \end{axis}
        \end{tikzpicture}
        \caption{$c = \SI{0.5}{\milli\gram\per\milli\liter}$.}
    \end{subfigure}
    \begin{subfigure}{0.33\textwidth}
        \centering
        \begin{tikzpicture}
        \begin{axis}[
            width=\textwidth, height=\textwidth,
            xlabel = Time {[s]}, ylabel = Measurement signal {[V]},
            xmin=0, xmax=9,
            y tick label style={
                /pgf/number format/.cd, fixed, fixed zerofill,
                precision=3, /tikz/.cd
            }]
            \input{addplot__Dilution01}
        \end{axis}
        \end{tikzpicture}
        \caption{$c = \SI{0.1}{\milli\gram\per\milli\liter}$.}
    \end{subfigure}
    \caption{Sensitivity test.
             Measured series of three pulses for different particle concentrations
             $c = \SIlist{10; 5.0; 1.0; 0.5; 0.1}{\milli\gram\per\milli\liter}$.}
    \label{fig: Results: Sensitivity test}
\end{figure*}

The measured signals for the six tested SPION suspension dilutions are shown in Fig.~\ref{fig: Results: Sensitivity test}. The three peaks that were expected due to the transmitted binary sequency with three equispaced ones can be identified in all measurements except for the one with the lowest SPION concentration $c = \SI{0.1}{\milli\gram\per\milli\liter}$. The amplitudes scale with the concentration and range from \SI{0.3}{\volt} ($c = \SI{10}{\milli\gram\per\milli\liter}$) to approx. \SI{0.02}{\volt} ($c = \SI{0.5}{\milli\gram\per\milli\liter}$).
For the considered setup the lowest concentration that allows safe detection of a bit is therefore determined as \SI{0.5}{\milli\gram\per\milli\liter}. Successful communication at lower concentrations may be possible but would require more robust equalization and coding schemes.

\section{Conclusion}
An optimized receiver for detection of superparamagnetic iron oxide nanoparticles in a molecular communication (MC) testbed was proposed. It was shown that the detector is capable of discriminating different particle concentrations that pass through the propagation channel, such that text messages and longer bit sequences can be effectively transmitted.

Compared to the previously used, commercial susceptometer, the proposed sensor device shows three advantages:
First, it can be used with different detector coils, particularly custom made ones, which can be manufactured according to the dimensions of the MC testbed in use.
Second, it provides an increased sample rate, which is currently limited by the software of the data acquisition card we used and which can be enhanced further by a custom made software.
And third, the reference, a second coil that the detuning of the measurement coil is compared to, is made available for the user. In this way, on the one hand, conventional single-ended signaling can be implemented by fixing a tube with the channel medium (destilled water in our case) without particles as reference. On the other hand, differential signaling is now possible, too, by using two tubes as propagation channel: In this case, the transmitter encodes information in concentration differences of SPIONs between the two tubes rather than by particle injections into a single tube. This would open up new possibilities for transmitter designs, e.\,g., by steering particles into one tube or the other by magnetic fields, which would not suffer from the mechanical limitations of pumps and, in combination with a closed-loop system, from the need of a reservoir and a sink for the nanoparticles.

Therefore, the next steps of research includes the realization of a fully differential signaling testbed for molecular communication. Furthermore, the proposed sensor device will be evaluated and compared to the commercial susceptometer as well as to alternative measurement approaches, particularly with respect to bit error rates for different measurement setups. Finally, an optimized receiver will allow to improve the testbed design and to investigate appropriate encoding and decoding strategies in a more detailed manner than it has been possible so far.

\begin{acks}
This work was supported in part by the Emerging Fields Initiative (EFI) of the Friedrich-Alexander-Universitat Erlangen-N{\"u}rnberg (FAU), the STAEDTLER-Stiftung, and the German Federal Ministry of Eduction and Research (BMBF), project MAMOKO.%, project Makroskopische Molekulare Kommunikation.
\end{acks}

\bibliographystyle{ACM-Reference-Format}
\bibliography{references}

\end{document}